\documentclass[aps,prd,twocolumn,showpacs,showkeys,preprintnumbers]{revtex4-2}
\usepackage{amsmath}
\usepackage{geometry}

\usepackage{graphicx}% Include figure files
\usepackage{dcolumn}% Align table columns on decimal point
\usepackage{bm}% bold math
\usepackage{hyperref}% add hypertext capabilities
\usepackage{multirow}
\begin{document}

\preprint{APS/123-QED}

\title{Cascade ($\Xi^0$) Baryon Spectroscopy in the Relativistic Framework of Independent Quark Model}% Force line breaks with \\

\author{Rameshri V. Patel}\author{Manan N. Shah}
\email{rameshri.patel1712@gmail.com}
 \affiliation{P D Patel institute of Applied Sciences, Charusat University, Anand 388421, Gujarat, India.}%Lines break automatically or can be forced with \\
\email{mnshah09@gmail.com (Corresponding Author)}
%\affiliation{P. D. Patel institute of Applied Sciences, Charusat University, Anand 388421, Gujarat, India.}
\date{\today}% It is always \today, today,

\begin{abstract}
The spectroscopy of $\Xi^0$ is performed within the relativistic framework of independent quark model. The equal mixture of scalar and vector components in the potential having Martin-like form is considered for the confinement. With the suitable potential parameters for $\Xi^0$, mass spectra for high radial and orbital excitation is calculated. The experimentally observed values of ground state magnetic moment, branching ratios and asymmetry parameters for radiative weak decays,  $\Xi^{0}\rightarrow\Lambda^{0} + \gamma^{0}$ \& $\Xi^{0}\rightarrow\Sigma^{0} + \gamma^{0}$ are obtained to validate the model. The spin parity of experimentally known resonances like $\Xi(1530)$, $\Xi(1820)$, \& $\Xi(2030)$ are confirmed through the Regge trajectories in $(J,M^2)$ plane. The spin pa+rity of $\Xi(1950)$, $\Xi(2130)$, \& $\Xi(2250)$
are predicted using those Regge trajectories. The radiative decay width and magnetic moment of first resonance is also predicted.
\end{abstract}

%Use showkeys class option if keyword
                              %display 
                              
\maketitle

%\tableofcontents

\section{Introduction}
In modern experimental facilities of the $21^{st}$ century, researchers are exploring the characteristics of heavy baryon resonances. However, there remains limited theoretical knowledge about the low-lying baryon resonances which are observed in experiments conducted during the $20^{th}$ century. Given the complexity of baryons, consist of three quarks and phenomenology emerges as the optimal approach to delve into the dynamics of quarks within these systems. It provides valuable insights into understanding the behavior of the strong force within such intricate configurations. 
One intriguing strange baryon is the $\Xi^0$, with approximately $10$ observed resonances ($\Xi(1530)$, $\Xi(1620)$, $\Xi(1690)$, $\Xi(1820)$, $\Xi(1950)$, $\Xi(2030)$, $\Xi(2120)$, $\Xi(2250)$, $\Xi(2370)$, $\Xi(2500)$) \cite{ParticleDataGroup:2022pth}.
Many of these resonances were identified in bubble chambers prior to the 1980s. However, our understanding of $\Xi$ resonances remains limited. This complexity arises due to several factors: firstly, $\Xi$ resonances can only be generated as part of a final state, making the analysis more intricate compared to direct formation. Secondly, their production cross sections are relatively small, typically in the range of a few microbarns. Lastly, the final states are characterized by topological complexity, posing challenges for study using electronic techniques \cite{ParticleDataGroup:2022pth}. For a thorough examination, we concentrate specifically on this cascade baryon, conducting spectroscopy to probe into its details.

On the theoretical side, K. Chao et al. used a nonrelativistic quark model \cite{Chao:1980em} then S. Capstick et al. used a relativized approach \cite{Capstick:1986ter}, both of which were based on one-gluon exchange. L. Y. Golzman et al. used the one-boson-exchange model \cite{Glozman:1997ag} while F. X. Lee et al. were focused on using QCD
sum rules \cite{Lee:2002jb}. Y. Oh was using the Skyrme model \cite{Oh:2007cr} for hyperon. Yan Chen and Bo-Qiang Ma studied the spectrum of light flavor baryons in a quark-model framework by taking into account the order $O(\alpha_s^2)$ hyperfine interactions due to two-gluon exchange between quarks \cite{Chen:2009de} which enabled them to produce masses till $\Xi \frac{7}{2}^+$. R. N. Faustov and V. O. Galkin in 2015, calculated high orbital and radial excitations of strange baryons by treating baryons as relativistic quark-diquark bound systems \cite{Faustov:2015eba}. Their primary assumption was that the two quarks with equal constituent masses form a diquark. C. Menpara and A. K. Rai employed the hypercentral constituent quark model with linear confining potential also a first order correction term to obtain the resonance masses and calculated states 1S-5S, 1P-4P, 1D-3D, 1F-3F, and 1G \cite{Menapara:2021dzi}. In very recent times, J. Oudichhya et al. extracted the relations between Regge slopes, intercepts, and baryon masses in the Regge phenomenology with quasi-linear Regge trajectories \cite{Oudichhya:2022off}. All of these treatments describe either the ground states or limited resonances while in this paper, our aim is to explain all the observed resonances and obtain their spin-parity using the independent quark model. 

The Independent Quark Model (IQM) was originally formulated by A. Kobushkin \cite{Kobushkin:1976fq} and P. Ferreira \cite{LealFerreira:1977gz} for the linear confinement of the quarks. In this approach, they considered that the individual quarks within a baryon follow a Dirac-type equation characterized by an average potential, defined within the center-of-mass of the Hadron. In subsequent advancements, scholars demonstrated that representing the average potential as an equal combination of scalar and vector components streamlines computations by transforming the single-quark Dirac equation into an effective Schrödinger equation \cite{Barik:1982nr, Barik:1985rm, Barik:1986gw}.
The investigation of quark confinement within a baryon can be conducted using the Martin-like potential, incorporating an equal mix of scalar and vector components. This potential has been applied in the relativistic context of the Independent Quark Model (IQM) for various mesons \cite{Shah:2014caa, Vinodkumar:2014afm, Shah:2014aly, Vinodkumar:2015blb, Shah:2014yma, Shah:2016mgq}. Given its favorable outcomes and its efficacy in predicting and validating experimental observations for mesons, recently, we adapted and enhanced this model to be applicable to various types of baryons \cite{Patel:2023wbs, Shah:2023mzg}. 

In this paper, we thoroughly describe the entire general methodology in section \ref{sec:2}, making it applicable to any type of baryon, in which we have outlined the process of solving Dirac equations for each quark within a baryon. It also involves determining spin-averaged masses and calculating contributions from spin-spin, spin-orbit, and tensor interactions. We computed static properties, such as the magnetic moment of this baryon which is already observed experimentally, providing a means to validate our model. Details of the procedure can be found in section \ref{sec:3}, along with the corresponding results. In the third section only, we describe the calculation of two decay widths for this baryon: radiative decay \ref{sec:3.1} and radiative weak decay \ref{sec:3.2}. Regge trajectories play a crucial role in phenomenological models by providing a framework for understanding and organizing experimental data on hadronic interactions, offering insights into the underlying dynamics of particles, and serving as a bridge between experimental observations and theoretical concepts. So, in the section \ref{sec:4}, we present our findings regarding the Regge trajectories in the $(J, M^2)$ plane.

\section{Methodology}\label{sec:2}
In these \cite{Shah:2014caa, Vinodkumar:2014afm, Shah:2014aly, Vinodkumar:2015blb, Shah:2014yma, Shah:2016mgq} studies, independent quark model has been applied primarily to meson systems. However, our contribution extends its application to encompass baryons, enabling spectroscopic analysis within this framework.
We explore constituent quarks within a hadron using the relativistic framework of the independent quark model. In this model, the properties of individual quarks are governed by a Dirac equation formulated in the hadron's rest frame. The potential in this equation showcases a Lorentz structure, featuring an even blend of scalar and vector components.

We propose that quarks within a hadron system experience independent motion within a flavor-independent central potential, characterized by a Martin-like form
\begin{equation}\label{potential}
    V(r) = \frac{(1+\gamma_0)}{2}(\Lambda r^{0.1}+V_0),
    \end{equation}
where, $\Lambda$ is the potential strength and $V_0$ is the depth of the potential. The Dirac equation for a quasi-independent quark in the center of mass frame has the form of 
\begin{equation}\label{dirac}
    \left[E^D_q - \boldsymbol{\hat{\alpha}} .\boldsymbol{\hat{p}} - \hat{\beta} m_q - V(r)\right]\psi_q(\vec{r}) = 0,
\end{equation}
where $E^D_q$ represents the Dirac energy of a quark, $m_q$ is the current quark mass and $\psi_q(\vec{r})$ is the four-component quark wave-function which is a spinor. As discussed in \cite{Greiner:2000cwh}, the solution of Eqn.(\ref{dirac}) can be expressed as 
\begin{equation}
    \psi_q(\vec{r}) = \begin{pmatrix}
    i g(r) \Omega_{jlm} \left(\frac{\boldsymbol{r}}{r}\right)\\
    -f(r) \Omega_{jl'm} \left(\frac{\boldsymbol{r}}{r}\right)
    \end{pmatrix}.
\end{equation}
Here, the spinor spherical harmonics $\Omega_{jlm}$ defined as given in Ref. \cite{Greiner:2000cwh},
\begin{equation}
    \Omega_{jlm} = \sum_{m',m_s} \big(l\frac{1}{2}j \rvert m'm_s m\big) Y_{lm'} \chi_{\frac{1}{2}m_s},
\end{equation}
with parity $\hat{P}_0 \Omega_{jlm} = (-1)^l \Omega_{jlm}$, $\chi_{\frac{1}{2}m_s}$ being eigenfunctions of $\boldsymbol{\hat{S}}^2$ $\&$ $\hat{S}_3$ and $Y_{lm}$ being the spherical harmonics. The radial parts of Dirac spinors follow
second-order ordinary differential equations.
    \begin{multline}\label{g}
        \frac{d^2g(r)}{dr^2} + \Bigg[(E^D_q+m_q)[E^D_q-m_q-V(r)]\\-\frac{k(k+1)}{r^2}\Bigg]g(r) = 0,
    \end{multline}
    \begin{multline}\label{f}
        \frac{d^2f(r)}{dr^2} + \bigg[(E^D_q+m_q)[E^D_q-m_q-V(r)]\\-\frac{k(k-1)}{r^2} \bigg]f(r) = 0,
    \end{multline}
where $k$ is the eigenvalue of the operator $\hat{k} = (1+\boldsymbol{\hat{L}}\cdot\boldsymbol{\hat{\sigma}})$ having the value, 
\begin{equation}
    k = \begin{cases} 
            -(l+1) = -\left(j+\frac{1}{2}\right) \hspace{0.7cm} for \hspace{0.1cm} j = l+\frac{1}{2}\\ \hspace{1.15cm}l = +\left(j+\frac{1}{2}\right) \hspace{0.7cm} for \hspace{0.1cm} j = l-\frac{1}{2}  
        \end{cases}
\end{equation}
These ODEs can be made equivalent to ODE obeyed by the reduced radial part of the Schrödinger wave function \cite{Barik:1982nr}

\begin{equation}\label{r}
        \frac{d^2R^{Sch}(r)}{dr^2} + \left[m_q(E^{Sch}_q-V(r))-\frac{l(l+1)}{r^2}\right]R^{Sch}(r) = 0.
\end{equation}
For the potential of Martin-like for Eqn.(\ref{potential}), we can define a dimensionless variable $\rho = \frac{r}{r_0}$ that will reduce these Eqns. (\ref{g}),(\ref{f}) \& (\ref{r}) to the equivalent form,
\begin{equation}\label{grho}
\frac{d^2g(\rho)}{d\rho^2} + \left[\epsilon^{D} - \rho^{0.1} - \frac{k(k+1)}{\rho^2}\right]g(\rho) = 0,
\end{equation}
\begin{equation}
\frac{d^2f(\rho)}{d\rho^2} + \left[\epsilon^{D} - \rho^{0.1} - \frac{k(k-1)}{\rho^2}\right]f(\rho) = 0,
\end{equation}
\begin{equation}
\frac{d^2R^{Sch}(\rho)}{d\rho^2} + \left[\epsilon^{Sch} - \rho^{0.1} - \frac{l(l+1)}{\rho^2}\right]R^{Sch}(\rho) =0.
\end{equation}
Here, 
\begin{equation}
\epsilon^D = (E^D_q - m_q - V_0)(m_q + E^D_q)^\frac{0.1}{2.1}\left(\frac{2}{\Lambda}\right)^\frac{2}{2.1}
\end{equation} 
and 
\begin{equation}
\epsilon^{Sch} = m_q\left(E^{Sch}_q - V_0\right)\left(m_q\right)^\frac{-2}{2.1}\left(\frac{1}{\Lambda}\right)^{\frac{2}{2.1}}.
\end{equation}
In the $Sch.$ case, $r_0 = (m_q \Lambda)^{\frac{-1}{2.1}}$ and $r_0 = \left[(m_q + E^D_q)\frac{\Lambda}{2}\right]^{\frac{-1}{2.1}}$ in the Dirac case \cite{Barik:1982nr}. The Schrödinger equation can be solved numerically using the code given in Ref.\cite{Lucha:1998} and the Dirac energies for the individual quarks can be found by equating $\epsilon^D$ to $\epsilon^{Sch}$. So this formalism can give the spin average masses of the three-body hadron system like a baryon (having constituent quark $q_1$, $q_2$, and $q_3$) as
\begin{equation}
M_{SA}^{q_1q_2q_3} = E^D_{q_1}+ E^D_{q_2} + E^D_{q_3} - E_{CM}.
\end{equation}
Where $E_{CM}$ is the parametric center of mass correction considered to remove the effects which come from the inability of the center of mass to remain invariant. We fit the potential parameters by equating the theoretical spin average mass with the experimental spin average mass of the $S$ wave, where the experimental spin average mass can be calculated as,
\begin{equation}
M_{SA} = \frac{\sum_J (2J+1)M_{nJ}}{\sum_J (2J+1)}.
\end{equation}
which take the form of $(M_{1/2}+2M_{3/2})/3$ for the $S$ waves of the baryon. With the fitted parameters, the spin average masses of the excited $S$ waves can also be calculated. Now we can remove the spin degeneracy by incorporating the spin spin interaction to the $M_{SA}$ by considering the total spin of the quark system as $\vec{J_{3q}}=\vec{J_1}+\vec{J_2}+\vec{J_3}$.
\begin{equation}\label{vjj}
\big<V^{jj}_{q_1q_2q_3}(r)\big> =  \sum_{i=1,i<k}^{i,k=3} \frac{\sigma \big<j_i.j_kJM \rvert \widehat{j_i}.\widehat{j_k}\rvert j_i.j_kJM\big>}{(E^D_{q_i}+m_{q_i})(E^D_{q_k}+m_{q_k})},
\end{equation}

which describes the interactions as the sum of the interaction of individual pairs of quarks. Here $\sigma$ is the $j-j$ coupling constant which can also be fitted using the experimental data. The fitted values of the potential parameters, center of mass correction, and the $j-j$ coupling constant for $\Xi^{0}$ baryon are given in the TABLE \ref{tab:uncertainities}. 

\begin{table}[!tbh]
\begingroup
\caption{Fitted parameters for the $\Xi^{0}$} \label{tab:uncertainities}
\setlength{\tabcolsep}{8pt}
\renewcommand{\arraystretch}{1.5}
\begin{tabular}{c c}
%\multicolumn{2}{c}{\textbf{TABLE I :} Fitted parameters for the $\Xi^{0}$}\\
\hline
\hline
Parameter & Value ( with $5\%$ variation)\\
\hline
Depth of the potential, $V_0$ & $-1.893 \pm 0.0946$ $GeV$  \\
Potential strength $\Lambda$  & $1.890 \pm 0.0945$ $GeV^{1.1}$ \\
Center of mass correction  $E_{CM}$  & $0.039 \pm 0.0019$ $GeV$ \\
$j-j$ coupling constant $\sigma$ & $0.067 \pm 0.0033$ $GeV^{3}$\\
\hline
\hline
\end{tabular}
\endgroup
\end{table}

To derive the masses of the $P$, $D$, \& $F$ states from the spin-averaged mass, we incorporate three interactions: spin-spin, spin-orbit, and tensor interactions. The spin-spin interaction term is defined in the equation (\ref{vjj}), while the spin-orbit and tensor interaction terms emerge as integral components of the confined one-gluon exchange potential \cite{Vinodkumar:1992wu}, which are also considered to be the sum of interactions between the pairs of quarks,
    % \begin{eqnarray}
    % V^{LS}_{q_1q_2q_3}(r) =  \frac{\alpha_s}{4} \sum_{i=1,i<k}^{i,k=3} \frac{N_{q_i}^2.  N_{q_k}^2}{(E^D_{q_i}+m_{q_i})(E^D_{q_k}+m_{q_k})} \frac{\lambda_i.\lambda_j}{2r}\\
    % \otimes [[r \times (\widehat{p}_{q_i}-\widehat{p}_{q_k}).(\widehat{\sigma}_{q_i}+\widehat{\sigma}_{q_k})].[(D'_0(r)+2D'_1(r))]\\
    % +[r \times (\widehatt{p}_{q_i}+\widehat{p}_{q_k}).(\widehat{\sigma}_{q_i}-\widehat{\sigma}_{q_k})].[(D'_0(r)-D'_1(r))]],
    % \end{eqnarray}

    \begin{multline}
V^{LS}_{q_1q_2q_3}(r) = \frac{\alpha_s}{4} \sum_{i=1}^{3} \sum_{k>i}^{3} \frac{N_{q_i}^2 N_{q_k}^2}{(E^D_{q_i} + m_{q_i})(E^D_{q_k} + m_{q_k})} \frac{\lambda_i \cdot \lambda_k}{2r} \\
\otimes \left[ \left[ \mathbf{r} \times (\widehat{\mathbf{p}}_{q_i} - \widehat{\mathbf{p}}_{q_k}) \cdot (\widehat{\sigma}_{q_i} + \widehat{\sigma}_{q_k}) \right] \left[ D'_0(r) + 2D'_1(r) \right] \right. \\
+ \left. \left[ \mathbf{r} \times (\widehat{\mathbf{p}}_{q_i} + \widehat{\mathbf{p}}_{q_k}) \cdot (\widehat{\sigma}_{q_i} - \widehat{\sigma}_{q_k}) \right] \left[ D'_0(r) - D'_1(r) \right] \right],
\end{multline}
    \begin{multline}
    V^{T}_{q_1q_2q_3}(r) =  -\frac{\alpha_s}{4} \sum_{i=1,i<k}^{i,k=3} \frac{N_{q_i}^2  N_{q_k}^2}{(E^D_{q_i}+m_{q_i})(E^D_{q_k}+m_{q_k})}\\ \otimes \lambda_i.\lambda_j \left(\left(\frac{D''_1(r)}{3}-\frac{D'_1(r)}{3r}\right)S_{{q_i}.{q_k}}\right).
    \end{multline}

    Where $\lambda_i .  \lambda_j$ represents the color factor of the baryon and  $S_{{q_i}.{q_k}} = [3(\sigma_{q_i}{\hat{r}})(\sigma_{q_k}{\hat{r}})-\sigma_{q_i}\sigma_{q_k}]$, the running coupling constant can be calculated as
\begin{eqnarray}
\alpha_s = \frac{\alpha_s(\mu_0)}{1+\frac{33-2n_f}{12\pi} \alpha_s(\mu_0) ln\left(\frac{E_{q1}^D+E_{q2}^D+E_{q3}^D}{\mu_0}\right)}.
\end{eqnarray} 
Where $\alpha_{s}(\mu_{0}=1GeV)=0.6$
is considered in the present study.
 We keep the parametric form of the confined gluon propagators $(D_0 \ \&  \ D_1)$ as it is mentioned in Ref
 .\cite{Vinodkumar:1992wu}
 and here prime represents the derivative with respect to r.
 \begin{equation} 
    D_0(r) = \left(\frac{\alpha_1}{r}+\alpha_2\right)exp\left(\frac{-r^2c_0^2}{2}\right)
    \end{equation}
    \begin{equation}
    D_1(r) = \frac{\gamma}{r}exp\left(\frac{-r^2c_1^2}{2}\right)
    \end{equation}
with $\alpha_1 = 10$, $\alpha_2 = 10$, $c_0 = 0.05$ $GeV$, $c_1 = 0.05$ $GeV$ and $\gamma = 10$. 
Consequently, upon deriving the wavefunction, we can determine the $N_{q_i}$, which is the normalisation constant for the individual quark wavefunction and the determination of $⟨\psi|V^{L S}|\psi⟩$ and $⟨\psi|V^T|\psi⟩$ becomes feasible by evaluating these quantities for all permutations of $q_1$, $q_2$, and $q_3$. The cumulative summation of these values provides the overall contribution for a specific state. The incorporation of spin-spin interaction contributions in addition to this total, yields the masses of corresponding $P$, $D$, and $F$ states.

\begin{table*}
%\fontsize{15pt}
%\resizebox{17.5cm}{!}
{
\begingroup
\caption{$S$ State masses (in $GeV$)} \label{tab:2}
\setlength{\tabcolsep}{3pt}
\renewcommand{\arraystretch}{2.5}
\begin{tabular}{l l l l l l l l l l l l}
%\multicolumn{12}{l}{\textbf{TABLE II :} $S$ State masses (in GeV)} \\
\hline
\hline
$nL$ & $J^{P}$ & State & $\big<V^{jj}_{q_1q_2q_3}\big>$ & \multicolumn{1}{p{2.5cm}}{\centering Our \\ predictions} & \multicolumn{1}{p{1.5cm}}{\centering Experimental \\ observations\\ \cite{ParticleDataGroup:2022pth}} & \multicolumn{1}{p{2cm}}{\centering Relativistic \\ quark-\\diquark \\ \cite{Faustov:2015eba}} & \multicolumn{1}{p{1.5cm}}{\centering two \\ gluon \\ exchange \\ \cite{Chen:2009de}} & \multicolumn{1}{p{1.5cm}}{\centering hypercentral \\ constituent \\ quark\\ model \\ \cite{Menapara:2021dzi}} & \multicolumn{1}{p{1.5cm}}{\centering Quark \\model \\ \cite{Capstick:1986ter}} & \multicolumn{1}{p{1.5cm}}{\centering Skyrme \\ model \\ \cite{Oh:2007cr}} & \multicolumn{1}{p{1.5cm}}{\centering Regge \\ phenom-\\enology \\ \cite{Oudichhya:2022off}} \\
\hline
$1S$ & $\frac{1}{2}^+$ & $1^2S_{\frac{1}{2}}$ & $-0.136$ & $1.324 \pm 0.038$ & $1.315$ & $1.330$ & $1.317$ & $1.322$ & $1.305$ & $1.318$ & $1.291$ \\
$1S$ & $\frac{3}{2}^+$ & $1^4S_{\frac{3}{2}}$ & $0.081$ & $1.541 \pm 0.038$ & $1.532$ & $1.518$ & $1.526$ & $1.531$ & $1.505$ & $1.539$ & $1.534$ \\

$2S$ & $\frac{1}{2}^+$ & $2^2S_{\frac{1}{2}}$ & $-0.082$ & $1.848 \pm 0.059$ & $-$ & $1.886$ & $1.750$ & $1.884$ & $1.840$ & $1.932$ & $1.886$ \\
$2S$ & $\frac{3}{2}^+$ & $2^4S_{\frac{3}{2}}$ & $0.049$ & $1.979 \pm 0.058$ & $-$ & $1.966$ & $1.952$ & $1.971$ & $2.045$ & $2.120$ & $1.966$ \\

$3S$ & $\frac{1}{2}^+$ & $3^2S_{\frac{1}{2}}$ & $-0.064$ & $2.143 \pm 0.071$ & $-$ & $2.367$ & $1.982$ & $2.361$ & $2.100$ & $-$ & $2.333$ \\
$3S$ & $\frac{3}{2}^+$ & $3^4S_{\frac{3}{2}}$ & $0.039$ & $2.245 \pm 0.070$ & $2.250$ & $2.421$ & $1.970$ & $2.457$ & $2.165$ & $-$ & $2.318$\\

$4S$ & $\frac{1}{2}^+$ & $4^2S_{\frac{1}{2}}$ & $-0.055$ & $2.351 \pm 0.080$ & $-$ & $-$ & $2.054$ & $2.935$ & $2.150$ & $-$ & $2.708$ \\
$4S$ & $\frac{3}{2}^+$ & $4^4S_{\frac{3}{2}}$ & $0.033$ & $2.439 \pm 0.079$ & $-$ & $-$ & $2.065$ & $3.029$ & $2.230$ & $-$ & $2.624$ \\

$5S$ & $\frac{1}{2}^+$ & $5^2S_{\frac{1}{2}}$ & $-0.049$ & $2.514 \pm 0.087$ & $-$ & $-$ & $2.107$ & $3.591$ & $2.345$ & $-$ & $3.036$ \\
$5S$ & $\frac{3}{2}^+$ & $5^4S_{\frac{3}{2}}$ & $0.029$ & $2.592 \pm 0.086$ & $-$ & $-$ & $2.114$ & $3.679$ & $-$ & $-$ & $2.897$ \\
\hline
\end{tabular}
\endgroup}
\end{table*}

\begin{table*}[tbh!]

{
\begingroup
\caption{$P$ State masses (in $GeV$)} \label{tab:3}
\setlength{\tabcolsep}{12pt}
\renewcommand{\arraystretch}{2}
\begin{tabular}{ c c c c c c c c c }
\hline
\hline
$n^{2S+1}L_J$ & $\big<V^{jj}_{q_1q_2q_3}\big>$& $\big<V_{q_1q_2q_3}^{L.S}\big>$ & $\big<V_{q_1q_2q_3}^T\big>$& Our & \cite{ParticleDataGroup:2022pth} & \cite{Faustov:2015eba}& \cite{Chen:2009de}&\cite{Menapara:2021dzi}\\
\hline
$1^2P_{\frac{1}{2}}$ & $-0.108$ & $-0.085$ & $-0.038$ & $1.567 \pm 0.053$ &$-$& $1.682$ & $1.772$ & $1.886$\\
$1^2P_{\frac{3}{2}}$ & $0.072$ & $-0.015$ & $0.001$ & $1.857 \pm 0.052$ & $1.823$& $1.764$ & $1.801$ & $1.871$\\
$1^4P_{\frac{1}{2}}$ & $-0.085$ & $-0.121$ & $-0.075$ & $1.518 \pm 0.053$ & $-$&$1.758$ & $1.894$ & $1.894$\\
$1^4P_{\frac{3}{2}}$ & $-0.126$ & $-0.050$ & $0.025$ & $1.647 \pm 0.052$ &$-$& $1.798$ & $1.918$ & $1.879$\\
$1^4P_{\frac{5}{2}}$ & $0.162$ & $0.044$ & $-0.005$ &$2.000 \pm 0.052$ &$-$& $1.853$ & $1.917$ & $1.859$\\
\hline
$2^2P_{\frac{1}{2}}$ & $-0.074$ & $-0.041$ & $-0.014$ & $1.985 \pm 0.067$ &$-$& $1.839$ & $1.926$ & $2.361$\\
$2^2P_{\frac{3}{2}}$ & $0.047$ & $-0.007$ & $0.001$ & $2.155 \pm 0.066$&$-$ & $1.904$ & $1.976$ & $2.337$\\
$2^4P_{\frac{1}{2}}$ & $-0.059$ & $-0.058$ & $-0.028$ & $1.969 \pm 0.067$ &$1.950$& $2.160$ & $-$ & $2.373$\\
$2^4P_{\frac{3}{2}}$ & $-0.079$ & $-0.024$ & $0.009$ & $2.021 \pm 0.067$ &$-$& $2.245$ & $-$ & $2.349$\\
$2^4P_{\frac{5}{2}}$ & $0.107$ & $0.021$ & $-0.002$ & $2.241 \pm 0.065$ &$-$&$2.333$ & $-$ & $2.318$\\
\hline
$3^2P_{\frac{1}{2}}$ & $-0.060$ & $-0.022$ & $-0.006$ & $2.245 \pm 0.077$ &$-$& $2.21$ & $-$ & $2.929$\\
$3^2P_{\frac{3}{2}}$ & $0.037$ & $-0.004$ & $0.000$ & $2.368 \pm 0.075$ &$-$& $2.350$ & $-$ & $2.894$\\
$3^4P_{\frac{1}{2}}$ & $-0.048$ & $-0.031$ & $-0.012$ & $2.242 \pm 0.076$ &$-$&$2.233$ & $-$ & $2.946$\\
$3^4P_{\frac{3}{2}}$ & $-0.063$ & $-0.013$ & $0.004$ & $2.262 \pm 0.077$ &$-$& $2.352$ & $-$ & $2.912$\\
$3^4P_{\frac{5}{2}}$ & $0.086$ & $0.011$ & $-0.001$ & $2.431 \pm 0.075$ &$-$& $-$ & $-$ & $2.865$\\
\hline
$4^2P_{\frac{1}{2}}$ & $-0.052$ & $-0.012$ & $-0.003$ & $2.436 \pm 0.084$ &$-$& $-$ & $-$ & $3.577$\\
$4^2P_{\frac{3}{2}}$ & $0.032$ & $-0.002$ & $0.000$ & $2.534 \pm 0.083$ &$-$& $-$ & $-$ & $3.532$\\
$4^4P_{\frac{1}{2}}$ & $-0.042$ & $-0.017$ & $-0.006$ & $2.438 \pm 0.084$ &$-$& $-$ & $-$ & $3.599$\\
$4^4P_{\frac{3}{2}}$ & $-0.054$ & $-0.007$ & $0.002$ & $2.445 \pm 0.084$ &$-$& $-$ & $-$ & $3.554$\\
$4^4P_{\frac{5}{2}}$ & $0.074$ & $0.006$ & $-0.000$ & $2.584 \pm 0.082$ &$-$& $-$ & $-$ & $3.494$\\
\hline
\end{tabular}
\endgroup}
\end{table*}

\begin{table*}
%\resizebox{8.5cm}{!}
{
\begingroup
\caption{$D$ State masses (in $GeV$)} \label{tab:4}
\setlength{\tabcolsep}{12pt}
\renewcommand{\arraystretch}{2}
\begin{tabular}{ c c c c c c c c c }
%\multicolumn{9}{c}{\textbf{TABLE IV :} $D$ State masses (in GeV)} \\
\hline
\hline
$n^{2S+1}L_J$ & $\big<V^{jj}_{q_1q_2q_3}\big>$& $\big<V^{L.S}_{q_1q_2q_3}\big>$ & $\big<V^{T}_{q_1q_2q_3}\big>$& Our &\cite{ParticleDataGroup:2022pth}&\cite{Faustov:2015eba} & \cite{Chen:2009de} & \cite{Menapara:2021dzi}\\
\hline
$1^2D_{\frac{3}{2}}$ & $-0.203$ & $-0.060$ & $-0.005$ & $1.755 \pm 0.063$ &$-$& $2.100$ & $1.970$ & $2.270$\\
$1^2D_{\frac{5}{2}}$ & $-0.041$ & $-0.002$ & $0.001$ & $1.981 \pm 0.062$ &$-$& $2.108$ & $1.959$ & $2.234$\\
$1^4D_{\frac{1}{2}}$ & $-0.074$ & $-0.126$ & $-0.023$ & $1.799 \pm 0.063$ &$-$& $1.993$ & $1.980$ & $2.310$\\
$1^4D_{\frac{3}{2}}$ & $-0.067$ & $-0.082$ & $-0.008$ & $1.865 \pm 0.063$ &$-$& $2.121$ & $2.065$ & $2.283$\\
$1^4D_{\frac{5}{2}}$ & $0.148$ & $-0.024$ & $0.002$ & $2.150 \pm 0.062$ &$-$& $2.147$ & $2.102$ & $2.247$\\
$1^4D_{\frac{7}{2}}$ & $0.229$ & $0.051$ & $-0.002$ & $2.302 \pm 0.061$ &$-$& $2.189$ & $2.074$ & $2.203$\\
\hline
$2^2D_{\frac{3}{2}}$ & $-0.134$ & $-0.031$ & $-0.002$ & $2.096 \pm 0.074$ &$2.120$& $2.144$ & $2.174$ & $2.819$\\
$2^2D_{\frac{5}{2}}$ & $-0.021$ & $-0.001$ & $0.000$ & $2.242 \pm 0.073$ &$-$& $2.213$ & $2.170$ & $2.771$\\
$2^4D_{\frac{1}{2}}$ & $-0.055$ & $-0.064$ & $-0.010$ & $2.134 \pm 0.073$ &$-$& $2.091$ & $2.107$ & $2.874$\\
$2^4D_{\frac{3}{2}}$ & $-0.050$ & $-0.042$ & $-0.003$ & $2.168 \pm 0.073$ &$-$& $2.149$ & $2.184$ & $2.838$\\
$2^4D_{\frac{5}{2}}$ & $0.102$ & $-0.012$ & $0.001$ & $2.354 \pm 0.072$ &$-$& $-$ & $2.205$ & $2.790$\\
$2^4D_{\frac{7}{2}}$ & $0.158$ & $0.026$ & $-0.001$ & $2.447 \pm 0.071$ &$-$& $-$ & $2.189$ & $2.729$\\
\hline
$3^2D_{\frac{3}{2}}$ & $-0.109$ & $-0.017$ & $-0.001$ & $2.318 \pm 0.082$ &$-$& $-$ & $2.252$ & $3.455$\\
$3^2D_{\frac{5}{2}}$ & $-0.015$ & $-0.000$ & $0.000$ & $2.430 \pm 0.081$ &$-$& $-$ & $2.239$ & $3.391$\\
$3^4D_{\frac{1}{2}}$ & $-0.046$ & $-0.036$ & $-0.005$ & $2.359 \pm 0.081$ &$-$& $2.367$ & $2.254$ & $3.527$\\
$3^4D_{\frac{3}{2}}$ & $-0.042$ & $-0.023$ & $-0.002$ & $2.379 \pm 0.081$ &$-$& $-$ & $-$ & $3.479$\\
$3^4D_{\frac{5}{2}}$ & $0.083$ & $-0.007$ & $0.000$ & $2.522 \pm 0.080$ &$-$& $-$ & $-$ & $3.415$\\
$3^4D_{\frac{7}{2}}$ & $0.130$ & $0.015$ & $-0.000$ & $2.590 \pm 0.079$ &$-$& $-$ & $-$ & $3.336$\\
\hline
$4^2D_{\frac{3}{2}}$ & $-0.094$ & $-0.009$ & $-0.001$ & $2.488 \pm 0.088$ &$-$& $-$ & $-$ & $-$\\
$4^2D_{\frac{5}{2}}$ & $-0.012$ & $-0.000$ & $0.000$ & $2.580 \pm 0.087$ &$-$& $-$ & $-$ & $-$\\
$4^4D_{\frac{1}{2}}$ & $-0.040$ & $-0.020$ & $-0.003$ & $2.530 \pm 0.088$ &$-$& $-$ & $-$ & $-$\\
$4^4D_{\frac{3}{2}}$ & $-0.037$ & $-0.013$ & $-0.001$ & $2.542 \pm 0.088$ &$-$& $-$ & $-$ & $-$\\
$4^4D_{\frac{5}{2}}$ & $0.072$ & $-0.004$ & $0.000$ & $2.662 \pm 0.086$ &$-$& $-$ & $-$ & $-$\\
$4^4D_{\frac{7}{2}}$ & $0.113$ & $0.008$ & $-0.000$ & $2.714 \pm 0.086$ &$-$& $-$ & $-$ & $-$\\
\hline
\end{tabular}
\endgroup}
\end{table*}

\begin{table*}
%\resizebox{8.5cm}{!}
{
\begingroup
\caption{$F$ State masses (in $GeV$)} \label{tab:5}
\setlength{\tabcolsep}{15pt}
\renewcommand{\arraystretch}{2}
\begin{tabular}{c c c c c c c c}
%\multicolumn{8}{c}{\textbf{TABLE V :} $F$ State masses (in GeV)} \\
\hline
\hline
$n^{2S+1}L_J$ & $\big<V^{jj}_{q_1q_2q_3}\big>$& $\big<V^{L.S}_{q_1q_2q_3}\big>$ & $\big<V^{T}_{q_1q_2q_3}\big>$& Our & \cite{ParticleDataGroup:2022pth}&\cite{Faustov:2015eba}  & \cite{Menapara:2021dzi}\\
\hline
$1^2F_{\frac{5}{2}}$ & $-0.113$ & $-0.046$ & $-0.001$ & $2.032 \pm 0.081$ &$2.025$& $2.411$ & $2.713$\\
$1^2F_{\frac{7}{2}}$ & $-0.344$ & $0.004$ & $0.000$ & $1.852 \pm 0.071$ &$-$& $2.460$  & $2.647$\\
$1^4F_{\frac{3}{2}}$ & $-0.395$ & $-0.100$ & $-0.008$ & $1.689 \pm 0.072$ &$-$& $2.252$  & $2.786$\\
$1^4F_{\frac{5}{2}}$ & $-0.538$ & $-0.061$ & $-0.002$ & $1.592 \pm 0.073$ &$-$& $-$  & $2.733$\\
$1^4F_{\frac{7}{2}}$ & $0.430$ & $-0.011$ & $0.001$ & $2.613 \pm 0.068$ &$-$& $2.474$ & $2.667$\\
$1^4F_{\frac{9}{2}}$ & $0.581$ & $0.048$ & $-0.001$ & $2.820 \pm 0.067$ &$-$& $2.502$  & $2.588$\\
\hline
$2^2F_{\frac{5}{2}}$ & $-0.072$ & $-0.023$ & $-0.001$ & $2.292 \pm 0.079$ &$-$& $-$ &  $3.333$\\
$2^2F_{\frac{7}{2}}$ & $-0.239$ & $0.002$ & $0.000$ & $2.151 \pm 0.080$ &$-$& $-$ &  $3.249$\\
$2^4F_{\frac{3}{2}}$ & $-0.289$ & $-0.051$ & $-0.004$ & $2.044 \pm 0.081$ &$-$& $-$ &  $3.426$\\
$2^4F_{\frac{5}{2}}$ & $-0.393$ & $-0.031$ & $-0.001$ & $1.963 \pm 0.082$ &$-$& $-$ &  $3.358$\\
$2^4F_{\frac{7}{2}}$ & $0.303$ & $-0.006$ & $0.001$ & $2.686 \pm 0.075$ &$-$& $-$ &  $3.274$\\
$2^4F_{\frac{9}{2}}$ & $0.412$ & $0.0244$ & $-0.000$ & $2.823 \pm 0.074$ &$-$& $-$ &  $3.173$\\
\hline
$3^2F_{\frac{5}{2}}$ & $-0.057$ & $-0.013$ & $-0.000$ & $2.472 \pm 0.086$ &$-$& $-$ &  $-$\\
$3^2F_{\frac{7}{2}}$ & $-0.197$ & $0.001$ & $0.000$ & $2.348 \pm 0.087$ &$-$& $-$ &  $-$\\
$3^4F_{\frac{3}{2}}$ & $-0.242$ & $-0.028$ & $-0.002$ & $2.271 \pm 0.088$ &$-$& $-$ & $-$\\
$3^4F_{\frac{5}{2}}$ & $-0.330$ & $-0.017$ & $-0.000$ & $2.196 \pm 0.089$ &$-$& $-$  & $-$\\
$3^4F_{\frac{7}{2}}$ & $0.251$ & $-0.003$ & $0.000$ & $2.792 \pm 0.082$ &$-$& $-$  & $-$\\
$3^4F_{\frac{9}{2}}$ & $0.342$ & $0.014$ & $-0.000$ & $2.898 \pm 0.081$ &$-$& $-$ & $-$\\
\hline
$4^2F_{\frac{5}{2}}$ & $-0.049$ & $-0.007$ & $-0.000$ & $2.616 \pm 0.091$ &$-$& $-$  & $-$\\
$4^2F_{\frac{7}{2}}$ & $-0.172$ & $0.001$ & $0.0000$ & $2.501 \pm 0.093$ &$-$& $-$ & $-$\\
$4^4F_{\frac{3}{2}}$ & $-0.214$ & $-0.016$ & $-0.001$ & $2.442 \pm 0.094$ &$-$& $-$  & $-$\\
$4^4F_{\frac{5}{2}}$ & $-0.292$ & $-0.010$ & $-0.000$ & $2.372 \pm 0.095$ &$-$& $-$  & $-$\\
$4^4F_{\frac{7}{2}}$ & $0.221$ & $-0.002$ & $0.000$ & $2.893 \pm 0.088$ &$-$& $-$  & $-$\\
$4^4F_{\frac{9}{2}}$ & $0.301$ & $0.008$ & $-0.000$ & $2.981 \pm 0.087$ &$-$ & $-$ & $-$\\
\hline
\end{tabular}
\endgroup}
\end{table*}

The sensitivity of the model can be studied by understanding the uncertainty of the fitted parameters given in TABLE \ref{tab:uncertainities}. To achieve that, we first investigated the spin average masses change by considering $5\%$ change in the $\Lambda$, we observe an average $\sim 18\%$ change in the $M_{SA}$. However, the overall uncertainty associated with $M_{SA}$ can be understood by changing all three parameters, i.e., $\Lambda$, $V_0$, \& $E_{CM}$. Here we have considered $5\%$ change in all the parameters leads to the overall change of $\sim 3\%$ in the $M_{SA}$. We have mentioned our predictions along with uncertainty associated with the masses ( by considering $5\%$ variation in $\Lambda$, $V_0$, $E_{CM}$, \& $\sigma$), Although the $5\%$ variation in $\sigma$ has a negligible impact, we have included it in our calculations but omitted it from the mass table. Our results are compared with corresponding experimental observations and other theoretical predictions for the masses in the $S$, $P$, $D$, and $F$ states, as shown in Tables \ref{tab:2}, \ref{tab:3}, \ref{tab:4}, and \ref{tab:5}, respectively.

\section{Magnetic moments and Decay properties}
\label{sec:3}
The magnetic moment of baryons is expressed in relation to its constituent quarks \cite{Patel:2007gx} as follows:
\begin{eqnarray}
    \mu_{B} = \sum_{q}<\phi_{sf}|\vec{\mu}_{qz}|\phi_{sf}>, %magnetic moment
    \end{eqnarray}
    where
    \begin{eqnarray}\label{m}
    \mu_{q} = \frac{e_{q}}{2 m_{q}}\sigma_{q}. %effective mass mag moment
    \end{eqnarray}
tHere, $e_q$ and $\sigma_q$ represent the charge and the spin of the quark, and $|\phi_{sf}>$ is spin-flavor wave function. Inside the baryon, the mass of quarks may undergo alterations as a result of their binding interactions with the other two quarks. To consider this impact of the bound state, we incorporate the bound state effect by substituting the mass parameter from equation (\ref{m}) with the introduction of an effective mass. The effective quark masses $m_{q}^{eff}$ in our model is defined as
\begin{eqnarray} 
    m_{q}^{eff} = E^D_q\left(1+\frac{<H>-E_{CM}}{\sum_q E^D_q}\right).
\end{eqnarray}
which follows the property of $M_J = \sum_{q=1}^{3} m_{q}^{eff}$. Our prediction and comparison with other different approaches are given in TABLE \ref{tab:6}.

\begin{table}[!tbh]
%\resizebox{8cm}{!}
{
\begingroup
\caption{Magnetic moments {(in $\mu_N$)}} \label{tab:6}
\setlength{\tabcolsep}{8pt}
\renewcommand{\arraystretch}{2}
\begin{tabular}{ c c c c c }
%\multicolumn{5}{c}{\textbf{TABLE VI :} Magnetic moments {(in $\mu_n$)}} \\
\hline
\hline
State & Our & \cite{ParticleDataGroup:2022pth} & \cite{Menapara:2021dzi} & \cite{Lee:2005ds}\\
\hline
$\Xi^{0} \frac{1}{2}^+$ & $-1.42$ & $-1.25\pm0.014$ & $-1.50$ & $-1.37$\\
$\Xi^{0} \frac{3}{2}^+$ & $0.15$ & $-$ & $0.766$ & $0.16$ \\
\hline
\end{tabular}
\endgroup}
\end{table}

\subsection{Radiative decay}
\label{sec:3.1}
Radiative decays of baryons offer enhanced insights into the intrinsic structure of baryons and their correlation with the mass of constituent quarks. The radiative decay width of light baryons, such as $\Xi^0$, is relatively modest when compared to heavy baryons. However, it is not insignificantly small. Given the distinctive status of the cascade as a light baryon, it is worthwhile to delve into the calculation of its radiative decay width. The expression for the electromagnetic radiative decay width can be formulated based on the radiative transition magnetic moment (in $\mu_N$) and photon energy ($q$ $=$ $M_{3/2}-M_{1/2}$) as \cite{Dey:1994qi,Thakkar:2010ij},
\begin{eqnarray}
\Gamma_R=\frac{q^3}{4\pi}\frac{2}{2J+1}\frac{e^2}{m_p^2}|\mu_{\frac{3}{2}^+\rightarrow\frac{1}{2}^+}|^2,
\end{eqnarray}
where the transition magnetic moment takes the form, 
\begin{multline}
\mu_{\frac{3}{2}^+\rightarrow\frac{1}{2}^+}=\sum_{i}\big<\phi_{sf}^{\frac{3}{2}^+}|\mu_i\cdot\vec{\sigma_i}|\phi_{sf}^{\frac{1}{2}^+}\big>\\=\frac{2\sqrt{2}}{3}(\mu_u-\mu_s).
\end{multline}

The key distinction in this transition magnetic moment lies in how we determine the magnetic moment of the quarks participating in this process. We obtain this by taking the geometric mean of their effective masses\cite{Thakkar:2010ij, Dhir:2009ax}, as illustrated in the equation below.
\begin{eqnarray}
    m_i^{eff}=\sqrt{m_{i(\frac{3}{2}^+)}^{eff} m_{i(\frac{1}{2}^+)}^{eff}}.
\end{eqnarray}
Our obtained results are discussed in the last section. 
\subsection{Radiative Weak decay}
\label{sec:3.2}
The most dominant decay of $\Xi^0$ is $\Xi^0\rightarrow\Lambda \pi^0$, but, the radiative weak decay is the second and it is observed experimentally. The calculation of decay widths of the transitions $\Xi^0\rightarrow\Lambda \gamma$ \& $\Xi^0\rightarrow\Sigma^0 \gamma$ will be helpful in verifying our model and its parameters. 
We calculate the decay width for the radiative weak decays of $\Xi^0$ using a joint description of weak radiative (WR) and nonleptonic (NL) hyperon decays (HD) in broken $SU(3)$. The two groups of decays are linked via $SU(2)_W$ spin symmetry and vector-meson dominance (VMD) \cite{Zenczykowski:2005cs}. The effective Lagrangian for weak radiative hyperon decay $B_i \rightarrow B_f \gamma$ is 
\begin{eqnarray}
    \Bar{u}_f i \sigma_{\mu\nu}(p_f-p_i)^\nu (C+D\gamma_5)u_iA^\mu,
\end{eqnarray}
where, $C$ \& $D$ are parity-conserving and parity-violating amplitudes. For this case, the decay is given by \cite{Zenczykowski:2005cs},
\begin{eqnarray}
\Gamma=\frac{1}{\pi}\left(\frac{m_i^2-m_f^2}{2m_i}\right)^3(|C|^2+|D|^2),  
\end{eqnarray}
and the asymmetry by \cite{Zenczykowski:2005cs}
\begin{eqnarray}
    \alpha = \frac{2 Re(C^*D)}{|C|^2+|D|^2}.
\end{eqnarray}
Here, parity-conserving amplitudes obtained in the ground-state baryon pole
model from the NLHD amplitudes via the $SU(2)_W$+VMD route, are given by \cite{Zenczykowski:2005cs},
\begin{multline}
C(B_i\rightarrow B_f\gamma)=\left(\frac{e}{g}\right)\frac{1}{(m_i+m_f)\sqrt{2}}\\ \times B(B_i\rightarrow B_f U^0)
\end{multline}
$B$ describe amplitudes for the emission of a linear superposition $U^0$ of virtual vector mesons $\rho^0, \omega, \phi,$ corresponding to a
photon and obtained by the $SU(6)_W$ symmetry from the
NLHD amplitudes and given by,
\begin{multline}
B(\Xi^0\rightarrow \Lambda^0U^0)=\bigg[\frac{-1}{\sqrt{3}}\left(3\frac{fp}{dp}-1\right)(\mu_{\Xi^0}-\mu_{\Lambda^0})\\-\left(\frac{fp}{dp}+1\right)
\mu_{\Sigma\Lambda}\bigg]\frac{N}{\mu_pD},
\end{multline}
and
\begin{multline}
B(\Xi^0 \rightarrow\Sigma^0U^0)=\bigg[\bigg(\frac{fp}{dp}+1\bigg)(\mu_{\Xi^{0}}-\mu_{\Sigma^{0}})\\+\frac{1}{\sqrt{3}}\left(3\frac{fp}{dp}-1\right)
\mu_{\Sigma\Lambda}\bigg]\frac{N}{\mu_p D}. 
\end{multline}
In this approach, the parity-violating WRHD amplitudes has the form,
\begin{multline}
D(B_i\rightarrow B_f\gamma^0)=\bigg(\frac{e}{g}\bigg)\frac{1}{(m_i-m_f)\sqrt{2}}\\ \times A(B_i\rightarrow B_fU^0)
\end{multline}
where amplitudes A are related by $SU(2)_W$ to the (vanishing in the soft-meson limit) correction terms in NLHD as,
\begin{eqnarray}
A(\Xi^0 \rightarrow \lambda^0U^0)=\frac{-2+\epsilon}{9\sqrt{3}}\frac{1-x}{1-x^2}b_R+\frac{1}{2\sqrt{3}}S_R
\end{eqnarray}
\begin{eqnarray}
A(\Xi^0 \rightarrow \Sigma^0U^0)=\frac{-1}{3}\frac{1+x}{1-x^2}b_R-\frac{5}{6}S_R
\end{eqnarray} 
Throughout, we have kept the notation and parameter values same as \cite{Zenczykowski:2005cs}. Our obtained branching ratios and asymmetry parameters are given in the TABLE \ref{tab:7}.

\begin{table}[!tbh]
%\resizebox{9cm}{!}
{
\begingroup
\caption{Radiative weak decay (Branching ratios and asymmetry parameters)} \label{tab:7}
\setlength{\tabcolsep}{2pt}
\renewcommand{\arraystretch}{1.5}
\begin{tabular}{c c c  }
%\multicolumn{3}{c}{\textbf{TABLE VII :} Radiative weak decay (Branching ratios and asymmetry parameters)} \\
\hline
\hline
Decay and Asymmetry & Our & \cite{ParticleDataGroup:2022pth}\\
\hline
$\Xi^{0}\rightarrow\Lambda^{0} + \gamma^{0}$ & $1.69\times 10^{-3}$ & $1.17\pm0.07\times10^{-3}$\\
$\Xi^{0}\rightarrow\Sigma^{0} + \gamma^{0} $ & $3.48\times10^{-3}$ & $3.33\pm0.10\times10^{-3}$  \\
$\alpha_{\Xi\Lambda^0\gamma}$ &$-0.6957$ &$-0.704\pm0.019\pm0.064$ \\
$\alpha_{\Xi\Sigma^0\gamma}$ &$-0.7707$&$-0.69\pm0.06$ \\
\hline
\end{tabular}
\endgroup}
\end{table}

\section{regge trajectories}
\label{sec:4}
\begin{figure}[hbt!]
    \includegraphics[width=0.5\textwidth]{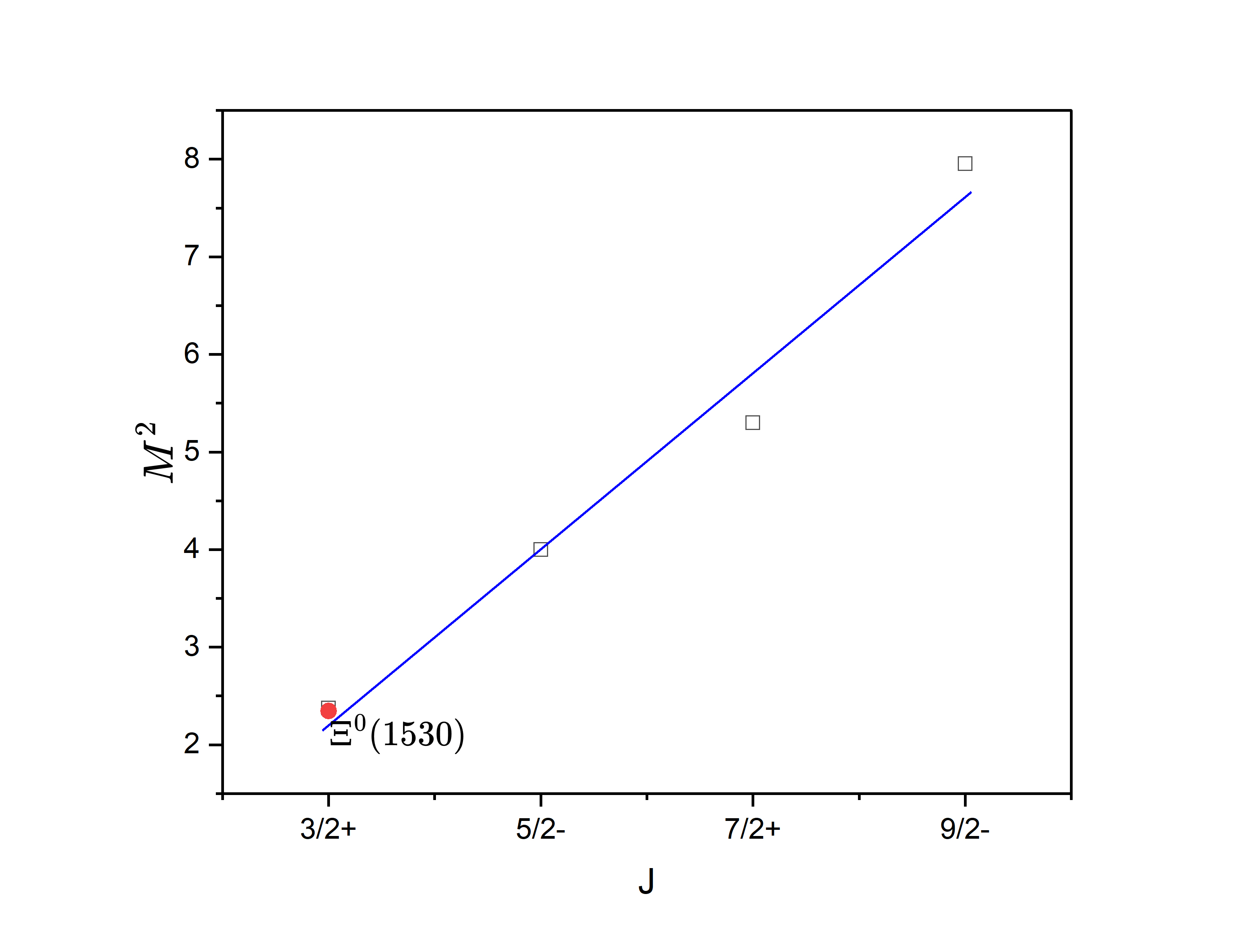}
    \caption{Regge trajectory for $1^4S_{\frac{3}{2}}$, $1^4P_{\frac{5}{2}}$, $1^4D_{\frac{7}{2}}$, \& $1^4F_{\frac{9}{2}}$ state masses. Red dot represents the experimental value of $\Xi(1530)$}
    \label{image1}
\end{figure}

\begin{figure}[hbt!]
    \includegraphics[width=0.5\textwidth]{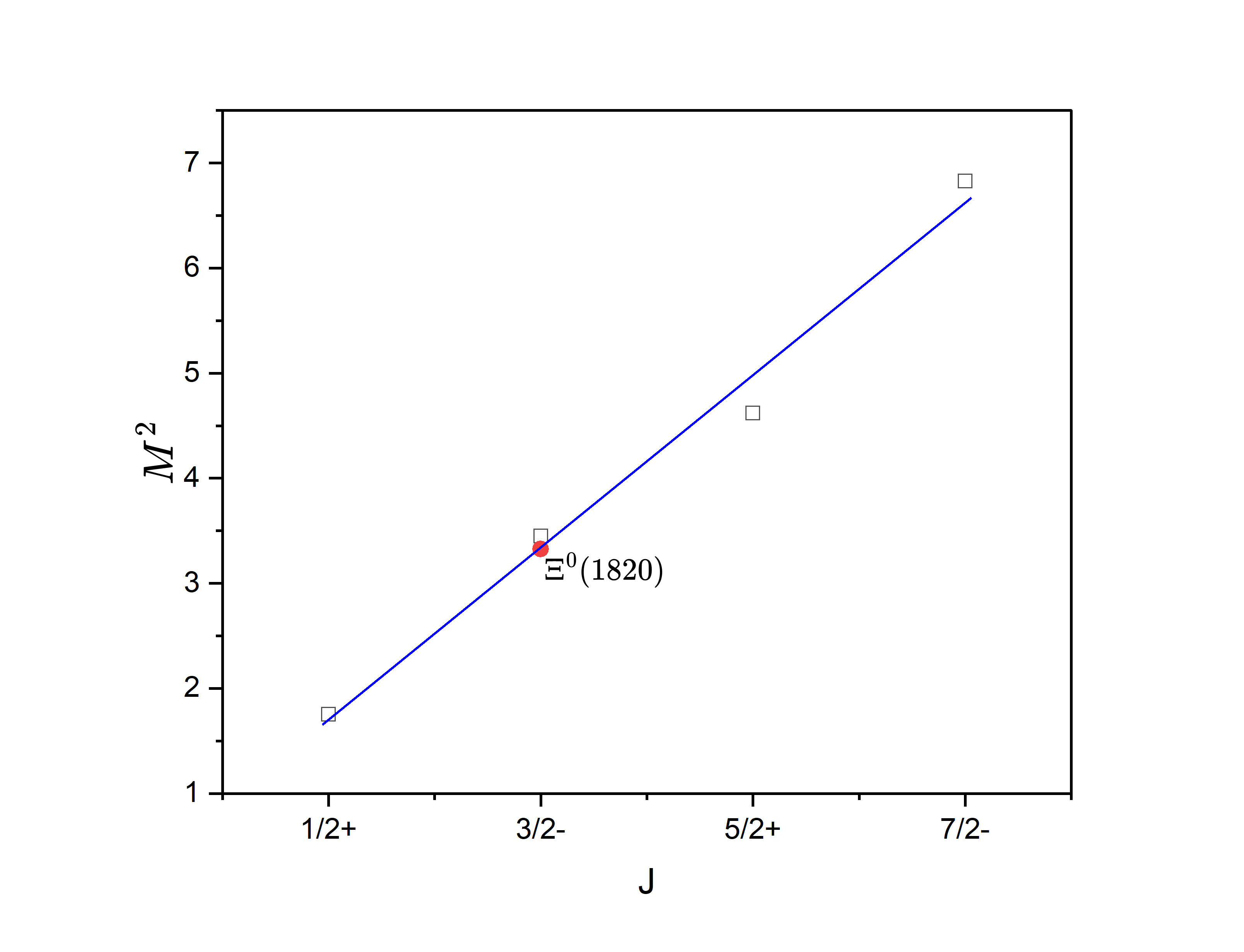}
    \caption{Regge trajectory for $1^2S_{\frac{1}{2}}$, $1^2P_{\frac{3}{2}}$, $1^4D_{\frac{5}{2}}$, \& $1^4F_{\frac{7}{2}}$ state masses. Red dot represents the experimental value of $\Xi(1820)$}
    \label{image2}
\end{figure}

\begin{figure}[hbt!]
    \includegraphics[width=0.5\textwidth]{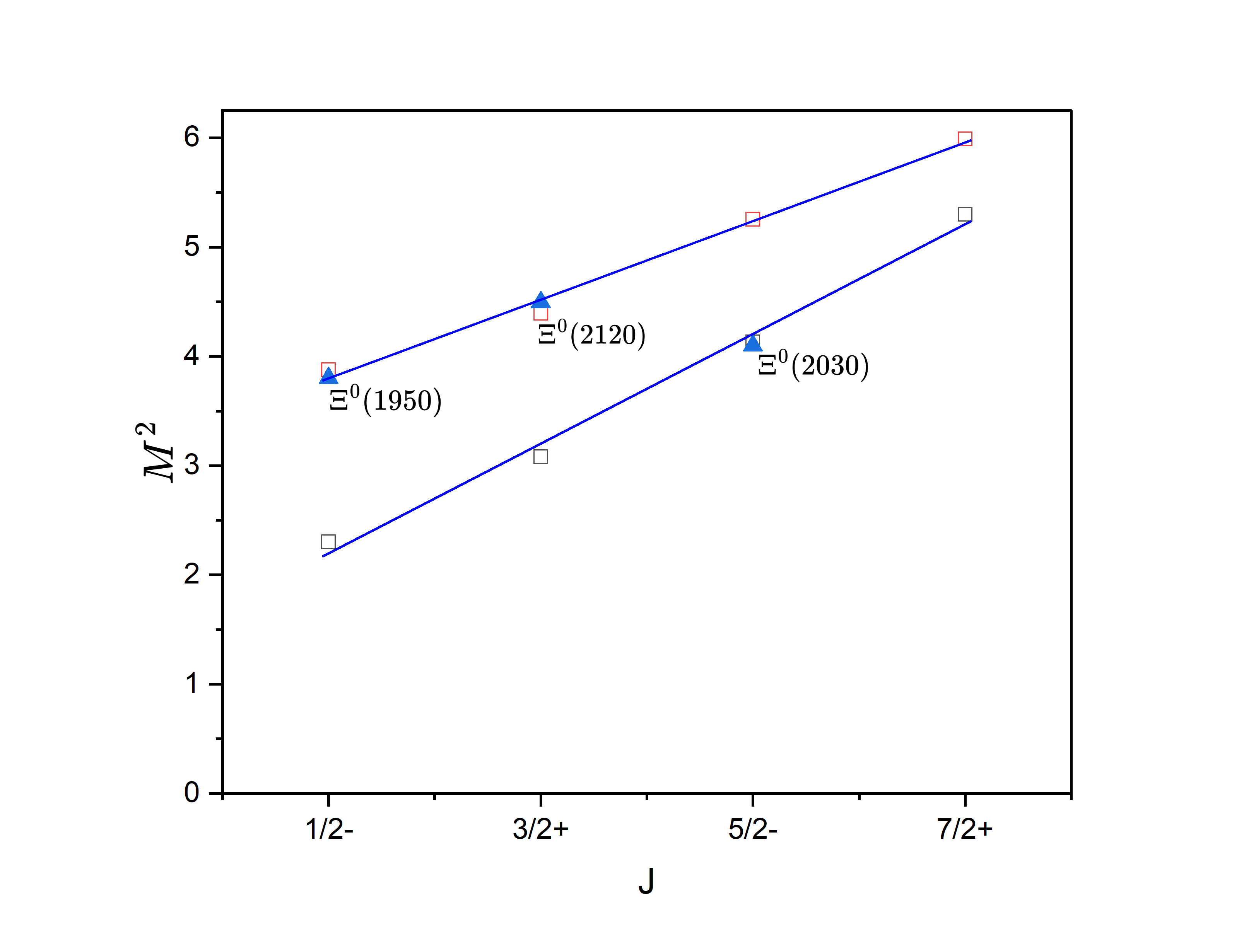}
    \caption{Regge trajectory for $1^4P_{\frac{1}{2}}$, $1^2D_{\frac{3}{2}}$, $1^2F_{\frac{5}{2}}$, \& $1^4D_{\frac{7}{2}}$ and $2^4P_{\frac{1}{2}}$, $2^2D_{\frac{3}{2}}$, $2^2F_{\frac{5}{2}}$, \& $2^4D_{\frac{7}{2}}$ state masses. Blue triangles represent the experimental values of $\Xi(1950)$, $\Xi(2120)$, \& $\Xi(2030)$.}
    \label{image3}
\end{figure}

\begin{figure}[hbt!]
    \includegraphics[width=0.5\textwidth]{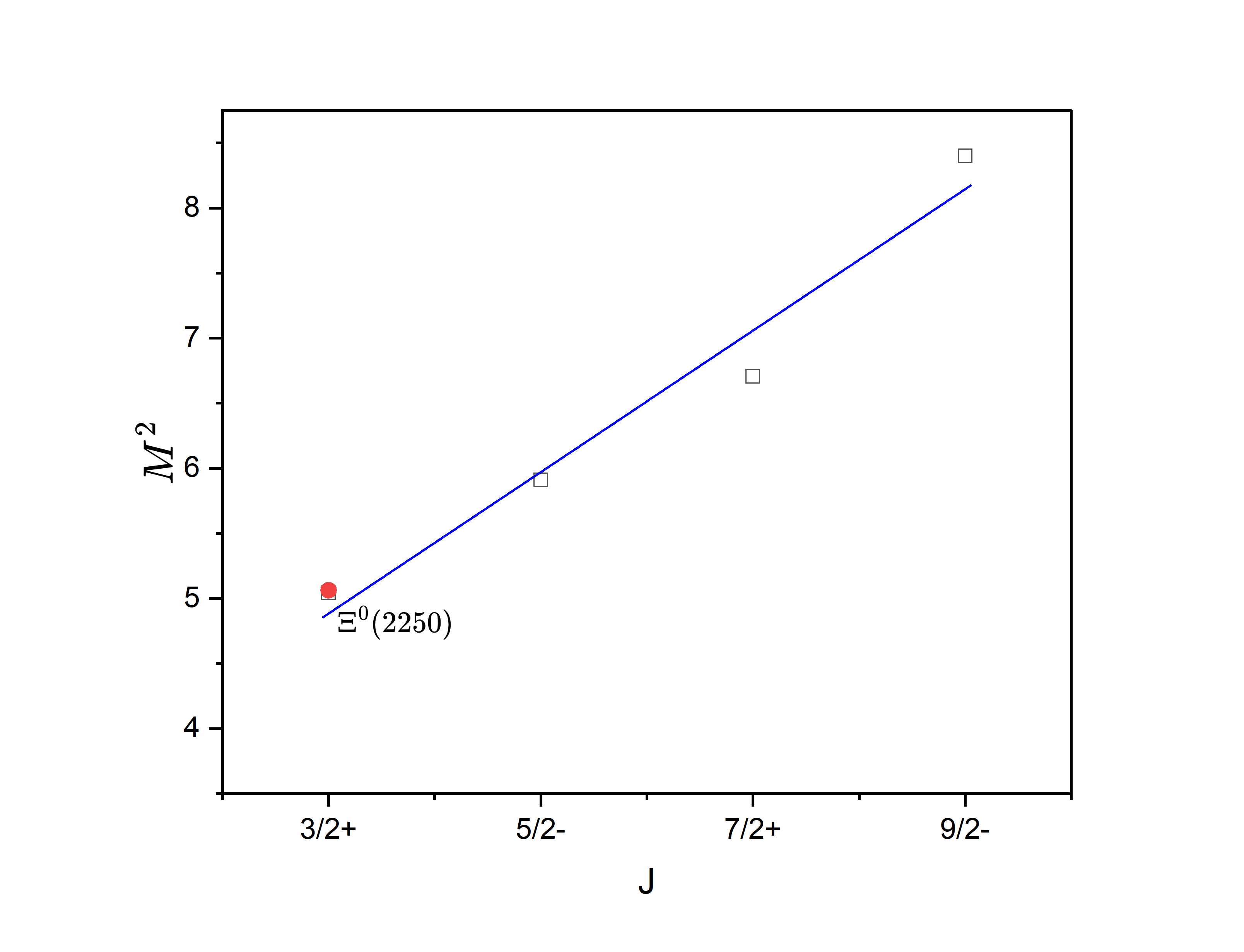}
    \caption{Regge trajectory for $3^4S_{\frac{3}{2}}$, $3^4P_{\frac{5}{2}}$, $3^4D_{\frac{7}{2}}$, \& $3^4F_{\frac{9}{2}}$ state masses. Red dot represents the experimental value of $\Xi(2250)$}
    \label{image4}
\end{figure}

Regge trajectories help in understanding the assignment of quantum numbers to hadronic states. This includes spin, parity, and other quantum numbers that characterize particles. Having determined the masses of orbitally and radially excited heavy baryons up to a high excitation numbers, we can construct the regge trajectories for this baryon in $(J,M^2)$ planes. We fit the linear relation
\begin{eqnarray}
    J=\alpha M^2+\alpha_0.
\end{eqnarray}
The plots are given in Fig. \ref{image1}, \ref{image2}, \ref{image3}, \& \ref{image4}; slopes and intercepts are given in TABLE \ref{tab:8}.

\begin{table}[!tbh]
%\resizebox{9cm}{!}
{
\begingroup
\caption{Slope($\alpha$) and intercept($\alpha_0$)} \label{tab:8}
\setlength{\tabcolsep}{12pt}
\renewcommand{\arraystretch}{1.5}
\begin{tabular}{  c c c   }
%\multicolumn{3}{c}{\textbf{TABLE VIII :} Slope($\alpha$) and intercept(\alpha_0) } \\
\hline
\hline
Baryon State &Slope & Intercept\\
\hline
$\Xi^{0} \frac{3}{2}^+$  & $0.541\pm0.060$ & $-0.154\pm0.321$\\
$\Xi^{0} \frac{1}{2}^+$  & $0.602\pm0.05$ & $-0.005\pm0.228$\\
$\Xi^{0} \frac{1}{2}^-$ & $0.887\pm0.120$ & $-3.280\pm0.797$\\
$\Xi^{0} \frac{1}{2}^-$  & $0.989\pm0.062$ & $-1.161\pm0.24$\\
$\Xi^{0} \frac{3}{2}^+$  & $-1.378\pm0.092$ & $-4.221\pm0.455$\\
\hline
\end{tabular}
\endgroup}
\end{table}

\section{Discussion and Conclusion}
\label{sec:5}
The plotted Regge trajectories and the masses mentioned in the tables \ref{tab:2}, \ref{tab:3}, \ref{tab:4}, \& \ref{tab:5}, suggest that the spin parity of $\Xi(1530)$, $\Xi(1820)$, $\Xi(1950)$, $\Xi(2030)$, $\Xi(2130)$, \& $\Xi(2250)$ could be $\frac{3}{2}^+$, $\frac{3}{2}^-$, $\frac{1}{2}^-$, $\frac{5}{2}^-$, $\frac{3}{2}^+$, \& $\frac{3}{2}^+$ respectively, which perfectly agrees with the experimental prediction of $\Xi(1530)$ to be $\frac{3}{2}^+$, $\Xi(1820)$ to be $\frac{3}{2}^-$ \cite{Biagi:1986vs} and $\Xi(2030)$ to be $\geq$$\frac{5}{2}$ 
 \cite{Amsterdam-CERN-Nijmegen-Oxford:1977bvi}. In the $n^{2S+1}L$ notation they are $1^4S$, $1^2P$, $2^4P$, $1^2F$, $2^2D$, \& $3^4S$ respectively. The linear behaviour of squared masses are observed in the $(J,M^2)$ plane. The calculated slopes and intercepts will be useful in verifying the additivity of inverse slopes, factorization of slopes, and additivity of intercepts when this approach will be applied to $\Sigma$ and $\Omega$ baryons.  

As it is widely recognized that the magnetic moments of the ground state of nearly all light flavour baryons have been established experimentally, our next logical step, following the determination of their masses, involves the computation of static property like magnetic moment. This enables us to assess and validate theoretical models beyond mere mass matching. In our study, we employed the effective constituent quark masses within our model to calculate the magnetic moments.
The obtained magnetic moment of the ground state to be $-1.42$$\mu_n$ is very close to the observed value of $-1.25\pm0.014$$\mu_n$. However, the magnetic moment of first resonance is obtained to be $0.15$$\mu_n$ that is in the range of order of predictions from hypercentral constituent quark model \cite{Menapara:2021dzi} and background field method \cite{Lee:2005ds}. 

Our investigation has yielded the radiative decay width to be $0.159 MeV$, in which the transition magnetic moment from ($\frac{3}{2}\rightarrow\frac{1}{2}$) is $1.94\mu_{N}$ which is close to that is obtained in other approaches like eﬀective mass and screened charge scheme \cite{Dhir:2009ax}, hypercentral constituent quark model with linear confining potential \cite{Menapara:2021dzi}, chiral constituent quark model \cite{Dahiya:2018ahb} and Lattice QCD \cite{Leinweber:1992pv}.    

The challenge of Weak Radiative Hyperon Decays (WRHD) has persisted for around six decades. It can be seen as a weak-interaction-related counterpart to the matter of baryon magnetic moments. However, unlike the baryon magnetic moments issue, which was well understood early in the development of the quark model, WRHD continues to generate significant controversy. Recent observation of negative asymmetry of $\Lambda \rightarrow n\gamma$ \cite{BESIII:2022rgl} confirms the Hara’s-theorem-satisfying nature of WRHD \cite{Zenczykowski:2020hmg}. 
The good observational data for weak radiative decay contains their branching ratios and the asymmetry parameters of both the possible decays of $\Xi^0$ i.e., $\Xi^{0}\rightarrow\Lambda^{0} + \gamma^{0}$ \& $\Xi^{0}\rightarrow\Sigma^{0} + \gamma^{0}$. We obtain those values using the joint description of weak radiative (WR) and nonleptonic (NL) hyperon decays (HD) $SU(2)_W$ spin symmetry and Hara’s-theorem-satisfying vector-meson dominance (VMD) \cite{Zenczykowski:2005cs}. The calculated values of branching ratios and asymmetries $\alpha_{\Xi\Lambda^0\gamma}$ \& $\alpha_{\Xi\Sigma^0\gamma}$ are in close proximity of experimentally observed values. 

Having seen the potential of our approach to predict the experimental data in the natural way with the fixation of model parameters $\lambda$, $V_0$, \& $\sigma$, we can apply this approach to many other baryons that has good number of experimental data available. Having the parameters for the baryons observed experimentally, we can correlate the parameters with the mass of the baryon which then be extrapolated to the baryons with less or no experimental data. This will enable us to predict the masses and other spectroscopic properties of those unobserved baryons which will be useful in resolving the future experimental uncertainties.

\begin{acknowledgments}
Ms. Rameshri Patel acknowledges the financial support from the University Grants Commission (UGC-India) under the Savitribai Jyotirao Phule Single Girl Child Fellowship (SJSGC) scheme, Ref. No.(F. No. 82-7/2022(SA-III)). We would like to express our gratitude to Prof. P.C. Vinodkumar for the insightful knowledge he imparted throughout this work.
\end{acknowledgments}


\end{document}